# Theory of morphological transformation of viral capsid shells during maturation process

O.V. Konevtsova,[*] V.L. Lorman,[†] and S.B. Rochal[*]

[*]Faculty of Physics, Southern Federal University, 5 Zorge str., 344090 Rostov-on-Don, Russia
[†]Laboratoire Charles Coulomb, UMR 5221 CNRS and Université de Montpellier, pl. E. Bataillon, 34095 Montpellier, France

In the frame of the Landau-Ginzburg formalism we propose a minimal phenomenological model for a morphological transformation in viral capsid shells. The transformation takes place during virus maturation process which renders virus infectious. The theory is illustrated on the example of the HK97 bacteriophage and viruses with similar morphological changes in the protective protein shell. The transformation is shown to be a structural phase transition driven by two order parameters. The first order parameter describes the isotropic expansion of the protein shell while the second one is responsible for the shape symmetry breaking and the resulting shell faceting. The group theory analysis and the resulting thermodynamic model make it possible to choose the parameter which discriminates between the icosahedral shell faceting often observed in viral capsids and the dodecahedral one observed in viruses of the Parvovirus family. Calculated phase diagram illustrates the discontinuous character of the virus morphological transformation and shows two qualitatively different paths of the transformation in a function of two main external thermodynamic parameters of the *in vitro* and *in vivo* experiments.

## I. INTRODUCTION

Self-assembly and shape transitions in biological nanostructures with non-trivial topology are characterized by unconventional properties and unusual pathways. In contrast to classical condensed media, bionanoassemblies undergo a multistep process which involves several physical transformations. Virus self-assembly is a typical example of the multistep process [1]. Viral capsid, the protein shell which protects viral genome, also passes through several successive steps of the assembly process. At the first step the so-called procapsid shell self-assembles from the aqueous solution of individual viral proteins. Capsid maturation is the final step of the virus self-assembly process. During maturation virus acquires the ability to infect its host cell. The process is characterized by a number of considerable correlated structure changes in the procapsid shell resulting in an infectious virion. One of the most interesting features of maturation process in a whole series of viruses is a morphological variation of the capsid shell resulting in a shape transition from the spherical to the faceted polyhedral geometry. It is worth noting that maturation in the host cell secretory pathway is often accompanied by irreversible biochemical events in capsid proteins including protein domain cleavage or neighboring protein crosslinking [2,3]. However, irreversible events usually take place when the structural modification leading to final protein positions and orientations in the shell is already achieved (see e.g. [4]). Thus, the reversible structural transition during maturation process can be considered as an independent physical phenomenon and treated in the frame of condensed matter physics.

To illustrate the notions of the theory developed in our work we choose the example of bacteriophage HK97 as well as other viruses with similar structural characteristics. HK97 is one of the viruses studied in detail by high-resolution structural methods which show eventful maturation scenario. HK97 is a dsDNA virus with the capsid protein shell which has rotational icosahedral symmetry group *I* and consists of 420 proteins distributed in seven 60-fold general positions of this group. Capsid structure satisfies to the well-known Caspar and Klug geometrical model [5] which limits the number of proteins constituting the shell to N=60T where the number of different positions T should take the value $T = h^2 + k^2 + hk$, with h and k being non-negative integers.

One of the main physical phenomena taking place during the HK97 bacteriophage maturation is the morphological transformation of the *spherical* procapsid into the *faceted* capsid accompanied by the increase in the shell volume [6,7,8]. Mean capsid diameter varies during the transformation to more than 20%. The protein shell becomes also much thinner, with more homogeneous shell thickness, and acquires a pronounced *icosahedral shape*. In the host cell pathway these morphological changes are induced by the ATP-dependent genome packaging into the pre-assembled procapsid protein shell, but in *in vitro* experiments similar structure and shape variation is realized in the absence of viral genomic DNA by using controlled pH decrease of the buffer [9]. Multistep maturation process with flattening transition at intermediate stages is typical for a whole series of viruses (e.g. P22 phage, Herpes Simplex Virus, etc.) which have structural organization similar to that of bacteriophage HK97 [9]. Universal features of this process represent the fundamental interest for both physics of nanoassemblies and physics of biological systems, and shed new light on principles of phase transitions in nanostructures with non-trivial topology.

Detailed study of different maturation steps for HK97 both in *in vivo* and *in vitro* experiments is still not complete. As it is often the case, the study of living systems by physical methods encounters certain experimental difficulties. The structures of the procapsid and matured capsid shells are usually determined by means of X-ray crystallography and high-resolution cryoelectron microscopy (cryoEM) [8,10]. However, the intermediate states of the *in vivo* maturation process are inaccessible by these techniques. Biochemical *in vitro* experiments study maturation dynamics in buffers with controlled pH level [11,12] and manage to distinguish several intermediate states with faceted shapes and one intermediate state with a spherical one. They are stabilized at different pH levels and characterized by successive increase in the shell volume. In spite of this progress the relation between pH variation in *in vitro* experiments and the *in vivo* capsid transformations induced by the genome packaging with the help of motor proteins [3] remains unclear. Experimental problems in virus maturation dynamics studies make of theoretical and numerical modeling an important tool for understanding physical phenomena at the origin of the procapsid shell transformation into the mature infectious virion.

Several theoretical approaches have been tested for capsid morphological transformation modeling. Structural changes during maturation have been described in [13] as a condensation of several low-frequency modes in a model system with the icosahedral symmetry. Two types of modes were considered to be responsible for the capsid faceting, the modes of the capsid isotropic expansion (or compression) and the modes of the pentamer displacements. HK97 virus maturation was also modelled [14] in the frame of the simplified Landau-Ginzburg theory of phase transitions. In contrast with the classical Landau-Ginzburg approach, the model proposed in [14] did not take into account the normal mode symmetry, thus reducing the order parameter to a simplified scalar physical quantity and disconnecting the structure from the free energy form. At the same time additional terms dependent on continuous derivatives of the order parameter with respect to variables on the shell surface were introduced in the model free energy, though the shell consists of only 420 particles. The intermediate structures of the maturation obtained in the frame of this approach are spatially inhomogeneous and their interpretation on the basis of available experimental data is not straightforward. The model [14] was then modified [15] by introducing methods of continuous elasticity theory but still remained disconnected from the mode symmetry.

The aim of the present work is to perform detailed group theory analysis of low-frequency modes responsible for the morphological changes, and to propose in the frame of the Landau-Ginzburg approach a minimal model of the procapsid-to-capsid transformation for the HK97 bacteriophage with the well-defined physical meaning of the free energy terms. The model developed in the paper is suitable not only for the considered HK97 virus but for a whole series of viruses demonstrating maturation process accompanied by capsid faceting and discontinuous volume jump.

## II. CRITICAL ORDER PARAMETERS RESPONSIBLE FOR THE MORPHOLOGICAL TRANSFORMATION

As it was shown previously in [13], structural changes during maturation in viral capsid shell of the HK97 bacteriophage and in a series of similar viruses involve at least two low-frequency modes. First of them induces isotropic capsid expansion, the second one is responsible for the shell faceting. Let us now discuss symmetry characteristics of the shell and their incidence on the modes responsible for the morphological changes during maturation process. For that aim we will first distinguish the symmetry of the protein density distribution from the symmetry of the shell shape, and, second, compare the symmetry of corresponding distributions in the procapsid and capsid states. Due to the asymmetry (and namely to the chirality) of coat proteins the protein density distribution of both procapsid and capsid shells have the same chiral symmetry group $I$ of all icosahedral rotations [5,16]. In contrast, the shape of the procapsid shell is spherical with a good accuracy while it becomes icosahedral in the faceted capsid state. The morphological transformation we are interested in is related mainly to this shape symmetry breaking. Thus, both modes responsible for the transformation are classified in the model as spanning irreducible representations of the SO(3) symmetry group of a chiral protein shell with the initial spherical shape.

All possible modes of the spherical shell transformation are collective displacement fields of material points of the spherical surface. In the most general form, corresponding displacement fields contain both radial and tangent components, tangent components being in turn separated into stretching and shear fields spanning different representations of the SO(3) group (for full mode classification and detailed study of the corresponding mechanical problem see [17]). Radial displacement modes bringing the main contribution to the morphological changes, in the minimal model we will take into account explicitly only the radial part of the displacement fields responsible for the procapsid-to-capsid shape variation. Note also, that the stretching component of the tangent displacement field spans the same irreducible representation of the SO(3) group as the radial component does, and thus, its amplitude is directly determined by the coupling with the radial component [17].

Any field of radial displacement of a spherical shell is usually expanded in scalar spherical harmonics

$$u_r(\theta,\phi) = \sum_{l=0}^{\infty} \sum_{l=-m}^{l=m} A_{l,m} Y_{l,m}(\theta,\phi) \qquad (1)$$

where $\theta$ and $\phi$ are the angles of a standard spherical coordinate system. Corresponding Cartesian vector **R'** of a point on a deformed sphere surface has the form:

$$\mathbf{R'} = \langle (u_r + R)\sin\theta\cos\phi, (u_r + R)\sin\theta\sin\phi, (u_r + R)\cos\theta \rangle, \qquad (2)$$

where R is the radius of the initial non-deformed sphere. A normal mode responsible for the shape symmetry breaking spans an irreducible representation of the SO(3) group labeled by a fixed value of the index (i.e. wave number) $l$.

Spherical harmonic $Y_{00}$ describes isotropic expansion (or compression) of the procapsid shall and plays the role of the first one-dimensional fully symmetrical order parameter of the minimal model developed in our approach. The second order parameter responsible for the shape variation from the spherical to the icosahedral one is not so simple, and needs much more detailed discussion. The experimentally observed capsid shape is characterized by the average shape which has full symmetry of an icosahedron $I_h$ in contrast with the microscopic symmetry of its chiral protein distribution which is invariant with respect to the rotational symmetry of an icosahedron $I$. Because of the shape invariance with respect to spatial inversion, the expansion (1) is limited to spherical harmonics with *even* wave numbers $l$ only. Furthermore, there are additional, much stronger rules which select possible index $l$ values for an irreducible mode, which can drive a transition from the spherical shape to the shape with full icosahedral symmetry $I_h$. The analysis based on the invariant theory (see Appendix) shows that these modes correspond to the functions $Y_{l,m}(\theta,\phi)$ with indices $l$ satisfying following selection rules:

$$l = 6i + 10j , \qquad (3)$$

where $i$ and $j$ are positive integers or zero. The sequence $L$ of the index $l$ values, where $l \in L$, allowed by (3) has the form : $L = (6, 10, 12, 16, 18, 20, 22, 24 ...)$. It selects spherical harmonics, and, consequently, symmetry breaking modes which can give a contribution to the icosahedral faceting of a viral capsid.

Explicit form of the displacement field $u_r(\theta,\phi)$ in the state with the icosahedral shape is obtained in terms of orthogonal functions $f_l^i(\theta,\varphi)$ with full icosahedral symmetry

$$u_r(\theta,\phi) = A_{0,0} + \sum_{l \in L} \sum_{i=1}^{i=n_t} D_{l,i} f_l^i(\theta,\phi) \qquad (4)$$

where $D_{l,i}$ are the amplitudes of the orthogonal icosahedral functions in the displacement field. The index $i$ ($i=1,...,n_t$) in the sum in (4) runs over all functions with the same fixed wave number $l$ value. The number of $n_t$ values is equal to the number of non-negative integer solutions $(i, j)$ of equation (3) for a given allowed $l$ value, and is quite limited. Consequently, the number of linearly independent icosahedral functions $f_l^i(\theta,\varphi)$ is also limited. According to Eq. (3) this number is equal to $n_t = 1$ for all $l<30$. Explicit form of the icosahedral functions $f_l^i(\theta,\varphi)$ with a given fixed $l$ value is easily obtained by averaging of spherical harmonics $Y_{lm}$ over the full icosahedral symmetry group:

$$f_l^i(\theta,\varphi) = 1/120 \sum_G Y_{lm}(\hat{g}(\theta,\varphi)) . \qquad (5)$$

Number $n_t$ can be obtained by a cumbersome but more accessible classical method widely used in condensed matter physics, especially in Raman and IR spectroscopies for active modes determination [18]. It involves the well-known character relation of the representations for the symmetry groups of the symmetric state and the state with the broken symmetry (with non-zero mode amplitude). For the procapsid spherical shell $2l+1$ harmonics $Y_{l,m}(\theta,\phi)$ with $m=-l, -l+1, \ldots l$ span one $(2l+1)$–dimensional irreducible representation of the SO(3) group. The icosahedral functions $f_l^i(\theta,\varphi)$ introduced above are mutually orthogonal basic functions of identity (fully symmetrical) representations of the icosahedral capsid shell symmetry group. They are linear combinations of $Y_{l,m}(\theta,\phi)$ with the given $l$ which are formed by the restriction of the SO(3) group to the icosahedral $I_h$ group. The number of identity representations $n_t$ is given by the character relation in the form [19]:

$$n_t(l) = 1/|G|\sum_G \xi_l(\hat{g}) \tag{6}$$

Where the sum runs over the elements $\hat{g}$ of the icosahedral symmetry group, number of elements is $|G| = 120$, and $\xi_l(\hat{g})$ are the characters of the SO(3) representation with the fixed $l$ value. Taking into account the fact that all functions with even wave number $l$ are invariant with respect to the spatial inversion we limit sum (6) to 60 rotational elements of the icosahedral symmetry group. After additional partition of the representation characters into conjugacy classes Eq. (6) takes the following form:

$$n_t(l) = \frac{1}{60}(2l+1+15\xi_l(\pi)+20\xi_l(2\pi/3)+12\xi_l(2\pi/5)+12\xi_l(4\pi/5)), \tag{7}$$

where $\xi_l(\alpha) = \sin((l+1/2)\alpha)/\sin(\alpha/2)$ is the explicit form of the character for a rotation to the angle α matrix.

The full set of harmonics in the displacement field (4) defines any surface with the icosahedral symmetry $I_h$. However, to describe the shape symmetry breaking it is sufficient to take into account in the displacement field expansion only minimal set of critical harmonics. According to the Landau theory principles, corresponding modes give a critical contribution to the capsid free energy variation in the vicinity of the morphologic transformation point where the amplitudes of other non-critical harmonics in (4) are negligibly small. Displacement field (4) limited to critical harmonics gives the main contribution to the spherical shape variation to the icosahedral one. The main result of the group theory analysis presented above is the conclusion that the simplest mode responsible for the capsid shape symmetry breaking corresponds to the minimal wave number value $l=6$ allowed by selection rules (3). Corresponding icosahedral function $f_6(\theta,\varphi)$ is the basis function of the (unique in this case) fully symmetric representation of the $I_h$ group resulting from the restriction of the irreducible representation of the SO(3) group with $l=6$.

In Cartesian coordinates the $f_6(x,y,z)$ function has the following form:

$$f_6(x,y,z) = (y^2 - z^2\tau^2)(z^2 - x^2\tau^2)(x^2 - y^2\tau^2) - 1/21(\sqrt{5}-2), \tag{8}$$

where $\tau = (\sqrt{5}-1)/2$ is the golden mean, and coordinates <x, y, z> of a point on a unit sphere are related in a standard way to $\theta$ and $\varphi$ variables of the spherical coordinate system. Explicit substitution of this function in the displacement field expression leads to the minimal displacement field which breaks the shape symmetry from the spherical to the icosahedral one. Note that two opposite signs of the $f_6(x,y,z)$ function amplitude $D_{6,i}$ lead to two different surfaces with the icosahedral symmetry. For one sign the displacement field results in a surface with the icosahedral shape, while for the opposite sign the surface acquires the dodecahedral shape (see Fig. 1). Experimentally, in the case of bacteriophage HK97 considered in our work the capsid shape induced by the displacement field is icosahedral (see Fig. 1,c) [20]. However, there exist a whole series of viruses which display dodecahedral capsid shape. An example of viruses of this type is given by the pathogenic human parvovirus B19 (Fig. 1,d) [21]. It is remarkable that the minimal model developed in our work catches this structural difference and describes it in a very simple way.

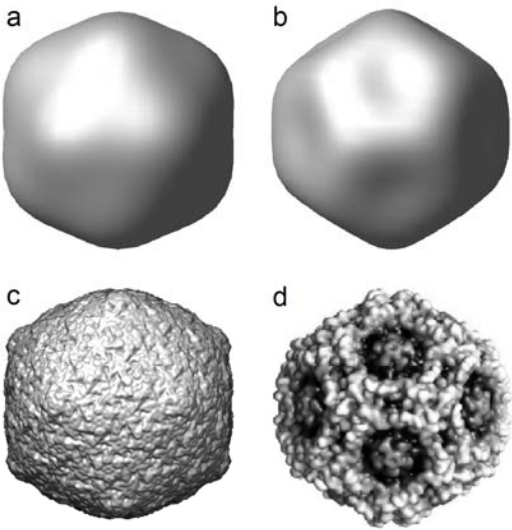

FIG. 1. (a-b): Spherical surface deformation by radial displacement fields proportional to the icosahedral function $f_6(\theta,\varphi)$. Amplitudes $D_{6,i}$ of the displacement fields in panels (a) and (b) have opposite signs. (c-d): Experimental realization of the capsid shapes induced by displacement fields (a) and (b). Viral capsid of the bacteriophage HK97 [20] with the icosahedral faceted shape (c), and viral capsid of the pathogenic human parvovirus B19 [21] (d) with the dodecahedral faceted shape.

### III.   FREE ENERGY OF THE PROCAPSID-TO-CAPSID MORPHOLOGICAL TRANSFORMATION

Symmetry and structure analysis performed in the previous Section shows that the main features of the shape variation during the maturation process in the HK97 bacteriophage and similar viruses are described by displacement fields with the minimal set of spherical harmonics with $l=0$ and $l=6$. These critical order parameters span two irreducible representations of the SO(3) symmetry group with the corresponding wave number values. The former field is responsible for the isotropic volume changes and the latter one breaks the shape symmetry from the spherical to the icosahedral one. Free energy of the transformation is dependent on the amplitude of the fully symmetric function with $l=0$, and 13 amplitudes $A_{6,m}$ of the harmonics with $l=6$. However, group symmetry analysis shows that these 13 amplitudes are linearly dependent in the state with the icosahedral symmetry. Consequently, all amplitudes $A_{6,m}$ in this state are proportional to the same value noted below as η. This "effective" order parameter value η defines the $D_{6,1}$ amplitude of the $f_6(\theta,\varphi)$ icosahedral function. The "effective" free energy depends only on η and on the variable describing isotropic volume variation. Value of the order parameter η obtained by the free energy minimization defines the degree of the capsid faceting, and the sign of η discriminate between icosahedral and dodecahedral shape of the resulting capsid. In addition, in what follows, instead of the $A_{00}$ amplitude of the fully symmetric harmonic $Y_{00}$ responsible for the isotropic expansion (or compression) we use explicitly capsid's volume variation ΔV as the corresponding variable in the free energy expansion. This choice is more suitable for physical interpretation of the results obtained in the frame of the developed approach. It is evident that the amplitude $A_{00}$ and the volume variation ΔV are related linearly. This justifies the change of variable proposed. Then, the minimal expansion of the free energy density describing the morphological transformation during maturation process takes the form:

$$F = A(d)\eta^2 + a_2\eta^3 + a_3\eta^4 - P\Delta V + b\Delta V^2 - |g|\Delta V \eta^2, \tag{9}$$

Phenomenological coefficient $A(d)$ in free energy density (9) is dependent on the capsid shell thickness $d$, which depends in turn on a whole series of biochemical parameters. Detailed discussion of the biochemical processes involved in the maturation and corresponding microscopic parameters are out of the scope of the present phenomenological work focused on the minimal model construction, and based on the full symmetry analysis of the morphological transformation problem. Here we limit our discussion to the experimental fact that decrease in the capsid shell thickness leads to the faceting transition. In the frame of the Landau approach developed in our work the stability limit for the spherical capsid shape corresponds to the coefficient $A(d)$ vanishing. But the faceting instability transition takes place even earlier, when $A(d)$ is still positive. Analysis of the morphological transformation thermodynamics described by free energy (9) shows that the considered shape transition is discontinuous. The procapsid undergoes first order faceting transition before full softening of the normal mode responsible for this shape transition takes place.

Discontinuous character of the faceting transition is directly related to the presence of the cubic term $a_2\eta^3$ in free energy density (9). It is worth noting that an invariant term which is cubic in a function of amplitudes of spherical harmonics, exists for any irreducible representation of the SO(3) group with even value of the wave number $l$. This fact was widely used previously in many different fields of physics. It is the case, for example, of nematic liquid crystal physics. The orientational ordering of rod-like molecules is described by the second-rank symmetric traceless tensor $Q_{ij}$, and the nematic order parameter spans the irreducible representation of the SO(3) group with $l=2$. It is evident that the determinant of the corresponding tensor is the cubic invariant of the irreducible representation. Due to even value of $l$ the same term is also invariant with respect to the full spherical symmetry group O(3). The same principles applied to other values of $l$ show that cubic invariant exists for all even wave numbers. Direct calculation of the cubic invariant for the irreducible representation with $l=6$ can be performed explicitly using properties of Clebsch-Gordan coefficients. Because of its cumbersome form we omit it in the present work. Note, that the cubic term in the free energy not only makes the transition discontinuous but also plays the crucial role in the choice between icosahedral and dodecahedral shapes of the resulting capsid shell.

Other terms in free energy density (9) have rather straightforward form and physical meaning. Because of the identical symmetry of the mode responsible for the isotropic volume change the free energy contains linear in Δ*V* term and the coupling term with the symmetry-breaking order parameter $\eta$ which is also linear in $\Delta V$. The term quadratic in $\Delta V$ with the positive coefficient $b>0$ ensures global stability of the isotropic expansion (or compression) mode. Fourth-degree term in order parameter η multiplied by the positive coefficient $a_3>0$ ensures the existence of a global minimum in the considered system. Coefficient *P* in free energy density (9) expresses the pressure difference between the inner and the outer regions of the capsid shell. Osmotic pressure difference created during the genome packaging into the capsid shell gives the main contribution to the corresponding term for *in vivo* maturation process. Negative sign of the coefficient *g* corresponds to the fact that both packaging-induced osmotic pressure and increase in capsid volume make the spherical shape less stable and favor morphological transformation to the faceted state. In a more complex model it is possible to take into account additional non-linearity of the free energy in a function of the isotropic expansion (or compression) mode represented here by the volume change variable $\Delta V$. Nonlinear in $\Delta V$ terms will favor intermediate states between spherical procapsid and icosahedral capsid states. But in the minimal model of the morphological transformation these terms lead to unjustified mathematical complications.

Minimization of the free energy functional with the density given by (9) leads to three possible solutions with different symmetries: i) $\eta=0$; ii) $\eta<0$; and iii) $\eta>0$, volume change $\Delta V$ being nonzero for all three states. Solutions ii) and iii) with opposite signs of the order parameter correspond to the shell with the icosahedral and the dodecahedral shape, respectively. They are usually called anti-isostructural states in the theory of phase transitions in condensed media [22]. Free energy density being non-linear in $\eta$ and quadratic in $\Delta V$, it is more convenient to minimize it first with respect to $\Delta V$, and substitute the solution in (9). The resulting "effective" free energy depends only on the symmetry breaking order parameter $\eta$ and its minimization leads to the same values of $\eta$ in the ordered states with the icosahedral symmetry of the shell

$$F(\eta) = (A(d) - |g|P/2b)\eta^2 + a_2\eta^3 + (a_3 - g^2/4b)\eta^4 - P^2/4b, \qquad (10)$$

Free energy density (10) has a simpler form and its behavior can be easily illustrated graphically. Its minimization is straightforward. The phase diagram of the model is presented in Fig. 2. Free energy (10) plots for several points typical for corresponding regions of the phase diagram are given in inserts a), b) and c) in Fig. 2. Phase diagram shows that the in-out pressure difference variation and the *A(d)* coefficient variation, induced by the shell thinning and the pH level decrease, contribute to the same part of the free energy.

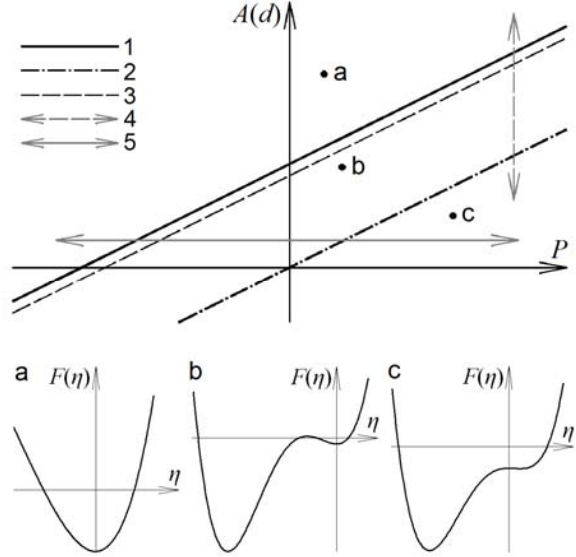

FIG. 2. Phase diagram of the free energy functional with density (9). Full line 1 and dash-dotted line 2 divide the phase diagram in qualitatively different regions: region of stability of the minimum with the spherical symmetry, region of stability of the minimum with the icosahedral symmetry, and the region where two minima with different symmetries coexist. Minimum with $\eta=0$ (spherical shape state without faceting) exists only in the region above line 2. The state with the icosahedral shape ($\eta<0$) appears below line 1. At dashed line 3 free energies of spherical and icosahedral faceted states are equalized. Below line 2 the minimum which corresponds to the spherical state disappears leaving the place for a weak metastable minimum with $\eta<0$. Inserts a), b) and c) show free energy density (10) as a function of the order parameter η in the typical points given by the same letters in the phase diagram. Two qualitatively different thermodynamic paths of the morphological transformation during maturation process are given by arrowed lines. Line 4 (downward) corresponds to the shell thinning induced by the pH level decrease. Line 5 (left to right) illustrates the result of the pressure difference increase during the *in vivo* genome packaging into the capsid.

The phase diagram in Fig. 2 was calculated for the case $a_2>0$. For negative values of this coefficient, free energy density (10) in inserts a), b) and c) is reflected with respect to the F axis and the ordered state with the icosahedral symmetry changes its shape from the icosahedral to the dodecahedral one. Corresponding lines in the phase diagram separate in this case the regions of stability of the spherical and the dodecahedral shell shapes. Dashed line 3 indicates in this case the equalization of the free energies of the spherical and the dodecahedral states.

Minimal model developed here allows us to calculate analytically the equation of line 3 where the free energies of the spherical and the icosahedral (or dodecahedral) states become equal:

$$A(d) = \frac{1}{2}\frac{4Pa_3bg - Pg^3 + 2a_2^2b^2}{(4a_3b - g^2)b}.$$

Another important quantitative characteristic which is also obtained analytically in the frame of the minimal model is the value of the volume jump at the transition from the spherical to the faceted state:

$$\Delta V = \frac{1}{8}\frac{g(3ba_2 + \sqrt{9a_2^2b^2 - 32a_3b^2A(d) + 16a_3bgP + 8g^2bA(d) - 4Pg^3})}{(4a_3b - g^2)^2b}$$

## IV. DISCUSSION AND CONCLUSION

The model developed in our work relates the faceting transition with two different external thermodynamic parameters of the system. First of them is the capsid shell thinning induced by the pH level decrease observed in *in vitro* experiments. The second one is connected to the *in vivo* genome packaging into the capsid shell with the help of motor proteins. Progressive capsid filling with the viral DNA leads to the pressure difference increase between the inner and outer regions of the shell. In the phase diagram of the model independent variation of these two external parameters corresponds to two different thermodynamic paths (shown by arrowed lines 4 and 5 in Fig. 2).

The capsid morphological transformation mechanism proposed here is consistent with classical works [23,24] on the capsid shell faceting based mainly on the continuous elasticity theory. These works have used the analogy between the faceting phenomenon and the longitudinal instability of disclinations in two-dimensional crystals. They have shown that in the locally hexagonal protein packing proposed as a model for viral capsid organization by Caspar and Klug [5] disclination instabilities in the vicinity of five-fold axes lead to the capsid faceting for viruses of sufficiently big size. The model proposed in [23] was based on the non-linear physics of thin elastic shells. Using continuous elasticity methods it elucidated the dependence of the viral shell faceting on the value of dimensionless Föppl–von Kármán (FvK) number $\gamma$, which characterizes the shell buckling instability. The FvK number is the combination $\gamma=YR^2/\kappa$, where $Y$ is the two-dimensional Young modulus of the shell, $\kappa$ is its bending rigidity, and $R$ is the mean radius of the capsid shell [23]. High bending rigidity favors smooth, practically spherical shell shape while low bending rigidity leads to the faceted shape. In the model developed in our work capsid shell thinning is shown to be one of the two important external parameters of the morphological transformation. It is evident that the shell thinning results in the decrease of its bending rigidity $\kappa$. The decrease in $\kappa$ leads to the shell faceting in a good accord with the predictions of [23].

In the recent work [15] the classical approach developed in [23,24] was submitted to a certain criticism. To propose an alternative theory the authors of [15] used the fact observed in *in vitro* experiments [3,8,25] that during the transition between two intermediate states EI and EII of the HK97 bacteriophage the protein hexamers constituting (together with pentamers) the viral capsid, change their shape from skewed to more regular one. Continuous elastic model developed in [15] proposed to relate the faceting transition not to the FvK number but to the hexamer shape change. However, X-ray crystallography and high-resolution cryoEM data reveal one more intermediate state of the HK97 maturation process which is characterized by the spherical capsid shape and the regular hexamer shape simultaneously [12]. This make experimental relevance of the model [15] not convincing enough.

We expect that further development of cryoEM technique will result in the near future in new high-resolution data on intermediate states of the HK97 morphological transformation. These additional data would constitute the basis for the further development of the minimal model proposed in the present work. The simplest extension would be the model which takes into account in (9) the terms of higher order in $\Delta V$. It is easy to see that by adding fourth–order terms in $\Delta V$ in (9) one obtains additional states which differ by their volume values. They might correspond to several spherical shells with different volumes observed in *in vitro* experiments [3,8,25].

In conclusion, let us note that the minimal model with a clear physical meaning of the free energy parameters proposed in the present work describes the viral capsid morphological transformation during maturation process for the HK97 bacteriophage as well as for a series of similar viruses. Underlying physical processes are driven by the order parameters spanning irreducible representation of the SO(3) symmetry group of the spherical shell constituted by asymmetric viral coat proteins. The morphological transformation during maturation process is understood as a sequence of phase transitions leading to the isotropic shell expansion and the symmetry-breaking faceting. The first fully symmetric order parameter is characterized by the shell volume change $\Delta V$. The second 13-dimensional order parameter responsible for the procapsid shape symmetry breaking describes explicitly icosahedral faceting of the viral shell. It spans the irreducible representation of the SO(3) group with $l=6$ and represents the linear combination of the spherical harmonics with $l=6$ invariant with respect to the icosahedral symmetry group $I_h$. In the state with the icosahedral symmetry the components of this order parameter depend on only one amplitude. In the minimal model of the transformation this fact is described by the effective one-dimensional order parameter $\eta$. The model is then reduced to the coupling between the fully-symmetric order parameter responsible for the capsid volume change and one-dimensional order parameter responsible for the shell faceting. It admits third-order term in $\eta$ in the free energy thus making the morphological transformation discontinuous. The third-order term sign discriminates between icosahedral and dodecahedral final shapes of the faceted capsid. The calculated phase diagram shows two qualitatively different paths of the transformation in a function of two main external thermodynamic parameters of the *in vitro* and *in vivo* experiments. The ensemble of the results obtained describes the experimentally observed physical phenomena which accompany maturation process in the HK97 bacteriophage and similar viruses.

**APPENDIX: DISPLACEMENT FIELDS WITH THE ICOSAHEDRAL SYMMETRY ON A SPHERICAL SURFACE**

Following analysis performed in terms of the invariant theory makes it possible to construct an arbitrary function with the icosahedral symmetry. It provides a justification for selection rules (3) which determine the possible wave number $l$ values associated with the order parameter responsible for the procapsid shape symmetry breaking from SO(3) to $I_h$.

We start with general properties of an arbitrary scalar function defined on a spherical surface (e.g. function of radial displacements of the capsid shell material points) and invariant with respect to the symmetry group $I_h$, which contains all symmetry operations of an icosahedron. Let us fix the point group orientation with respect to the coordinate frame and choose Cartesian axes *x*, *y* and *z* along the two-fold symmetry axes of an icosahedron. This choice allows us to express the invariants of the full icosahedral group in a simple way. Another property of the group helps us to construct so-called integrity basis constituted by the generators of the ring of invariant polynomials. Any function with the icosahedral symmetry can then be expanded in series of the finite number of invariant polynomials constituting the integrity basis. Point group $I_h$ belongs to the class of simple mathematical objects called groups generated by reflections. For the groups of this type the number of invariants in the integrity basis is equal to the dimension of its vector representation (i.e. to the dimension of the space in the considered case) and the product of degrees of basis invariants is equal to the number of the group elements $|G|$. Consequently, the integrity basis of the $I_h$ group contains only three following invariant polynomials with the degrees 2, 6 and 10, respectively:

$$J_0 = x^2 + y^2 + z^2, J_1 = \prod_{i=1}^{6} \mathbf{q}_i \mathbf{r}, J_2 = \prod_{i=1}^{10} \mathbf{p}_i \mathbf{r} , \qquad (11)$$

here $\mathbf{r} = <x, y, z>$; vectors $\mathbf{q}_i$ and $\mathbf{p}_i$ are parallel to the 5-fold and 3-fold icosahedral axes, respectively. From the geometric point of view each term in the products in the expressions for the invariant polynomials $J_1$ and $J_2$ is equivalent to the equation of a plane perpendicular to the 5-fold and 3-fold axes.

According to the well-known presentation of spherical harmonics as homogeneous polynomial functions any icosahedral function $f_l$ (given by Eq. (5)) with the fixed wave number value $l$ can be expressed as a homogeneous polynomial in $\mathbf{r} = <x, y, z>$ of the degree $l$. Taking unit radius-vector length we express $\mathbf{r}$ in spherical coordinates as:
$$x = \sin\theta\cos\varphi, y = \sin\theta\sin\varphi, z = \cos\theta.$$

On the unit sphere the invariant $J_0$ acquires constant value $J_0 = 1$. Then, any function on a spherical surface invariant with respect to the icosahedral symmetry group $I_h$ can be presented in the form:
$$f_l(\theta, \phi) = A_0 + A_1 J_1 + A_2 J_2 + A_3 J_1^2 + A_4 J_1 J_2 + A_5 J_1^3 + A_6 J_2^2 ... \quad (12)$$

The degree $l$ of the $f_l$ function is equal the degree (in a function of the radius-vector components) of the last terms in expansion (12). The number of terms of the degree $l$ in (12) is the number of possible integer linear combinations of numbers 6 and 10 equal to $l$. Consequently, for any homogeneous function of the degree $l$ invariant with respect to the $I_h$ group, the number $l$ satisfies to the condition $l=6i+10j$, where $i$ and $j$ are non-negative integers.

Due to the function irreducibility, coefficients $A_i$ multiplying the terms with the degrees smaller than $l$ in (12) are univocally defined by the orthogonality of function (12) to basic functions of other irreducible representations with the wave numbers $l'<l$. Following orthogonality relations hold for a given irreducible function $F$:
$$\iint F(\theta,\phi) Y_{l'm} \sin\theta d\phi d\theta = 0, \text{for all } l' < l \quad (13)$$

In practice the number of equations in system (13) is much higher than the number of unknown coefficients in function (12). However, the additional equations are either linearly equivalent or become identity.

## Acknowledgments

S.R. and O.K. acknowledge financial support by the RFBR grant 13-02-12085 ofi_m. (Russia). V.L. acknowledges financial support by the Laboratory of Excellence NUMEV (France), partial support by National Science Foundation Grant No. PHYS-1066293 and the hospitality of the Aspen Center for Physics.